\documentclass[12pt]{article}
\usepackage{graphicx}
\usepackage{amsmath}
\usepackage{caption}

\usepackage{times}

\title{The space group classification of topological band insulators}
\author
{Robert-Jan Slager$^{1}$, Andrej Mesaros$^2$, Vladimir Juri\v ci\' c$^{1\ast}$  and Jan Zaanen$^1$\\
\\
\normalsize{$^1$}{Instituut-Lorentz for Theoretical Physics, Universiteit Leiden}\\
\normalsize{P.O. Box 9506, 2300 RA Leiden, The Netherlands}\\
\normalsize{$^2$}{Department of Physics, Boston College}\\
\normalsize{Chestnut Hill, MA 02467, USA}\\
\normalsize{$^\ast$To whom correspondence should be addressed:
Vladimir Juri\v ci\' c; E-mail: juricic@lorentz.leidenuniv.nl.} }
\date{}

\begin{document}


\baselineskip24pt


\maketitle
\noindent
{\bf Topological band insulators (TBIs) are bulk insulating materials which feature topologically protected metallic states on their boundary.
The existing classification departs from time-reversal symmetry, but the role of the crystal lattice symmetries in the physics of these topological states remained elusive.
Here we provide the classification of TBIs protected not only by time-reversal, but also by crystalline symmetries. We find three broad classes of topological states: (a) $\Gamma$ states robust against general time-reversal invariant perturbations; (b) Translationally-active states protected from elastic scattering, but susceptible to topological crystalline disorder; (c) Valley topological insulators sensitive to the effects of non-topological and crystalline disorder. These three classes give rise to 18 different two-dimensional, and, at least 70 three-dimensional TBIs, opening up a route for the systematic search for new types of TBIs.}

\newpage
Topological phases of free fermionic matter are in general characterized by an insulating gap in the bulk and protected gapless modes on the boundary of the system\cite{hasan-kane-review,qi-zhang-review}. Integer quantum Hall states represent first examples of topologically protected phases in absence of any symmetries with the topological invariant directly related to the measured Hall conductance\cite{tknn1982}. Recently, it became understood that even in the presence of fundamental symmetries such as time-reversal, topologically protected states of matter can, in principle, exist. In particular, it has been shown that time-reversal invariant (TRI) insulators in two dimensions (2D)\cite{kane2005} and three dimensions (3D) \cite{moore2007,fukane2006,fukane2007a}  are characterized by  $Z_2$ topological invariants which pertain to the existence of the gapless boundary modes robust against time-reversal preserving perturbations, and may host Majorana quasiparticles\cite{fukaneMajorana2008}, as well as provide the condensed-matter realization of the theta-vacuum\cite{Zhang-Theta2010}. The theoretical prediction \cite{bernevigzhang2006,zhang2009} and experimental realization of the $Z_2$-invariant topological band insulators\cite{koning2006,hsieh2008,hsieh2009,xia2009,chen2009} gave a crucial boost in the understanding of these exotic phases of matter which culminated in the general classification of  topological insulators and superconductors based on time-reversal symmetry (TRS) and particle-hole symmetry (PHS) within the so-called tenfold periodic table \cite{schnyder2008,schnyderNJP2010, kitaev2009}.

The role of the crystal lattice in this classification is to provide a unit cell in the momentum space, the Brillouin zone (BZ), topologically equivalent to the $d$-dimensional torus, over which the electronic Bloch wavefunctions are defined. The tenfold classification follows then assuming that all the unitary symmetries of the corresponding Bloch Hamiltonian have been exhausted and therefore the only remaining symmetries are, according to the Wigner's theorem, antiunitary TRS and PHS. In three dimensions (3D), however, by considering a $Z_2$ TBI as a stack of two-dimensional (2D) ones, thus assuming a  layered 3D lattice, additional three ``weak'' invariants associated with the discrete translation symmetry have been found\cite{moore2007},
which characterize these states, and may be probed by lattice dislocations\cite{ranNatPhys2009}.
On the other hand, it has been realized  that topological states protected just by lattice symmetries, such as inversion, can exist in 3D \cite{fukane2007b,teofukane2008,fu2011,fuNatComm2012,hughesprodanbernevigPRB2011,turnerPRB2012}. Additionally, it has been recently found that a state protected both by TRS and the lattice $C_4$ rotational symmetry susceptible to the lattice dislocations can be possible in 2D\cite{juricic2012}. However, in general, the role of the space group symmetries in the physics of topological states remained elusive.

We here provide the complete classification of TBIs protected not only by TRS, but also by space group lattice symmetries. As a starting point, we depart from the construction by
Fu and Kane \cite{fukane2006,fukane2007a,fukane2007b} to compute the $Z_2$ invariant in terms of the matrix of overlaps
\begin{equation}\label{eq::w}
w_{mn}=\langle u_{m}(-\mathbf{k})|\vartheta | u_{n}(\mathbf{k})\rangle,
\end{equation}
where $\vartheta$ is the time-reversal operator and $|u_{n}(\mathbf{k})\rangle$ is the n-th occupied Bloch wavefunction. The quantities of central significance are
\begin{equation}\label{eq::pfaffian}
\delta_i=\frac{\sqrt{\text{det}[w({\Gamma}_{i})]}}{\text{Pf}[w({\Gamma}_{i})]}
\end{equation}
defined at the points ${\Gamma}_{i}$ in the BZ where the Hamiltonian commutes with the time-reversal operator. Since the matrix $w$ is antisymmetric at the points ${\Gamma}_i$, the Pfaffian is defined at these points and $\text{det}[w({\Gamma}_i)]=(\text{Pf}[w({\Gamma}_i)])^2$. The topological $Z_2$ invariant, $\nu$, is then given by $(-1)^\nu=\prod_{{\Gamma}_i}\delta_i$, and its non-triviality implies a topological obstruction for defining the wavefunctions through the entire BZ with an even number of band inversions.
Notice that the evaluation of the topological invariant in terms of the signs of the Pfaffian does not depend on the dimensionality but only on the fact that the Hamiltonian possesses TRS which, due to the vanishing of the Chern number, guarantees the existence of globally defined wavefunctions throughout the BZ.

First, notice that the set of the points ${\Gamma}_i$ at which the Hamiltonian commutes with the time-reversal operator is fixed by the space group of the lattice, see Table \ref{2dTable}. Second, we choose  the overall phase of the Bloch wavefunctions so that a unique phase, which we dub the ``$\Gamma$'' phase, has $\delta_{{\Gamma}}=-1$ at the ${\Gamma}$-point in the BZ and $\delta_i=1$ at all the other high symmetry points. A crucial observation is that \emph{the distribution} of signs of the Pfaffian, $\delta_i$, at the points $\Gamma_i$, and not only their product, encodes the additional topological structure. To show this, we first consider how the matrix of overlaps transforms under a lattice symmetry operation represented by a unitary operator $U$
\begin{equation}\label{eq::symmetryrelation}
w_{mn}(\mathbf{k})=\langle u_{m}(-\mathbf{k})|\vartheta | u_{n}(\mathbf{k})\rangle=\langle u_{m}(-U\mathbf{k})|U\vartheta U^{\dagger} | u_{n}(U\mathbf{k})\rangle=w_{mn}(U\mathbf{k}).
\end{equation}
As a consequence, when some of these high symmetry points are related by point-group symmetry of the lattice, their signs of the Pfaffian have to be equal.
Therefore, it is sufficient to consider a subset, $\Gamma_a$, of representative, inequivalent high symmetry points that are also \emph{not related} by any symmetry.
This leads to the following rule that allows for determination of all the topological phases given the space group and the corresponding high symmetry points, ${\Gamma}_i$: each phase is obtained by selecting a single representative high-symmetry point $\Gamma_a$ and setting $\delta_{{\Gamma_a}}=-1$, which automatically sets $\delta_{{\Gamma_b}}=-1$ at all the high-symmetry points $\Gamma_b$ related by point group symmetry to $\Gamma_a$. Such phases are separated by a topological quantum phase transition that involves bulk bandgap closing which changes the values of $\delta_i$'s.

Let us illustrate this simple classification principle by an elementary example. We start with the $\Gamma$ phase on a square lattice, $\delta_{{\Gamma}}=-1$, and $\delta_{{X}}=\delta_{{Y}}=\delta_{{M}}=1$ where $X$, $Y$, and $M$ are the time-reversal invariant (TRI) momenta in the BZ. By applying our rule, we immediately see that, besides the $\Gamma$ phase, we obtain an ``$M$'' phase with $\delta_{{M}}=-1$, and $\delta_{{\Gamma}}=\delta_{{X}}=\delta_{{Y}}=1$ (Table \ref{2dTable}). This phase is disconnected from the $\Gamma$ phase through a topological quantum phase transition with the bandgap closing at the ${X}$ and the ${Y}$ points. This phase is protected by TRS and is also susceptible to dislocations\cite{juricic2012}, and represents an example of a ``translationally-active'' phase.
Furthermore, since the $X$ and the $Y$ points are related by a $C_4$ rotation, there can exist a phase with $\delta_{X}=\delta_Y=-1$, and $\delta_\Gamma=\delta_M=1$.
The product of the $\delta_i$'s at all TRI momenta then yields the trivial $Z_2$ invariant, $\nu=0$. However, $C_4$ rotational symmetry protects this phase, since it pins the band inversions at the ${X}$ and ${Y}$ points.  This phase represents a ``valley'' (or ``crystalline''\cite{fu2011}) insulator -- a phase trivial tenfold way-wise but protected by the lattice symmetries. Indeed, this new phase, which we dub the ``$X$-$Y$'' phase, is realized in an extension of the $M$-$B$ model for a quantum spin Hall insulator with the next-nearest neighbor hopping, as shown in Fig.\ \ref{figurephasesquare}, and responds nontrivially to dislocations; see Supplementary Information for the technical details.
When $C_4$ rotational symmetry is reduced to $C_2$, the ${X}$ and ${Y}$ points are no longer related by symmetry, and therefore the symmetry constraint on the $\delta_{{X}}$ and $\delta_{{Y}}$ is no longer present. We then expect the $X\text{-}Y$ phase to be unstable, and to yield instead nontrivial phases with $\delta_{{X}}=-1$ or $\delta_{{Y}}=-1$, and $\delta_i=1$ at all other TRI momenta. Our calculations, indeed, confirm this within the M-B tight-binding model, as shown in Supplemetary Information A.  In general, an even number of TRI momenta related by symmetry yield a valley phase, protected by crystal symmetry while having $\nu=0$.

Let us now elaborate on the role of the space group of the underlying lattice in this classification, since this symmetry group defines the relation between the high symmetry points. The difference in phases found on rectangular and rhombic lattices serves as a clear illustration. Both these lattices have $D_2$ point-group symmetry, but different wallpaper groups (space groups in 2D).  The rhombic case has two inequivalent TRI momenta related by point group symmetry and hence a valley phase, see Table \ref{2dTable} and Supplementary Information, Sec. A. On the other hand, in the rectangular case all $D_2$ symmetry operations map any TRI momentum to its equivalent, thus no valley phase is possible. From Table \ref{2dTable} it is seen that in 2D the phases, as related to space groups, cluster in Bravais lattice classes, with an exception: the hexagonal lattice. We will see that this clustering is less generic in 3D. In turn, the primitive Bravais hexagonal (triangular) lattice ($p6mm$) is invariant under the $C_6$ rotational symmetry around a lattice site, as opposed to the non-primitive hexagonal lattice ($p3m1$) realized in graphene.
Based on our rule, we conclude that in the latter case only the $\Gamma$ phase is possible, which is in fact realized in the Kane-Mele model\cite{kane2005a}. In contrast, on the former lattice (triangular), the points $K_+$ and $K_-$ are related by a $C_6$ symmetry and thus each of these points becomes TRI. The number of TRI momenta is increased, ultimately yielding a possibility of additional translationally-active and valley phases, as shown in Table \ref{2dTable}. These phases are realized within the $M\text{-}B$ tight-binding model (Supplementary Information, Sec. A), and their robustness against disorder is shown in Supplementary Information, Sec. B.

The above rule allows us to completely classify and index the topological phases: the last entry in Table \ref{2dTable}. The set of BZ high-symmetry points $\Gamma_i$ at which there is band inversion, i.e., $\delta_{\Gamma_i}=-1$, is invariant under the operations of a subgroup of the lattice space group. This symmetry subgroup therefore protects and labels the topological phase. The other element in this indexing is the protection by TRS (T), existing when $Z_2$ invariant $\nu=1$, giving, for instance $T-p4mm$ as the $\Gamma$ phase on the square lattice. When the protecting symmetries coincide between phases, we explicitly label $\Gamma_i$ (lower index), as, e.g.,  for $T-p2m_X$, $T-p2m_Y$ and $T-p2m_M$ phases on the rectangular lattice.
This leads to the list of topological phases in 2D presented in Table \ref{2dTable} which gives 18 distinct topological phases.
As our general result, there are two additional broad classes of topological states protected by TRS or crystalline symmetries,
besides the class of states robust against general TRS perturbations ($\Gamma$-states): translationally-active states protected both by TRS and lattice symmetry, responding to dislocations, and valley insulators which are tenfold-way-wise trivial but protected by space group symmetry and also susceptible to dislocations.

Our procedure can be applied in the same way in 3D but it becomes more involved given the 230 space groups and the large number of high-symmetry points. We find at least 70 different phases\cite{us-inpreparation}. Here we will illustrate these matters for a number of simple crystal structures (Table \ref{3dTable}) which include those of TBIs that are of present empirical relevance\cite{hsieh2008,hsieh2009,xia2009,chen2009}. To illustrate matters, consider the primitive cubic lattice (Table \ref{3dTable}) with the familiar eight TRI points (Fig. \ref{fig:cubic}A). Crucially, the points $(X,Y,Z)$  are related by a three-fold rotation, as well as the points $(X',Y',M)$. Consequently, we obtain four TRS protected phases. We notice that this is quite different from the indexing procedure introduced by Moore and Balents \cite{moore2007}. For instance, our $T-pm{\bar 3}m$ ($\Gamma$) and $T-p3(4)_R$ ($R$) phases correspond with their $(1;0,0,0)$ and $(1;1,1,1)$ indices, respectively. Their latter two indices would also correspond with the $T-p3(4)_M$ and $T-p3(4)_X$ phases, respectively. The other possibilities in their classification are either coincident with our four TBIs, or represent a 3D phase not protected by crystal symmetries due to implicit dimensional reduction (e.g. layered 3D lattice); see Figs.\ \ref{fig:cubic}A,B.

The power of the space group classification becomes further manifest for non-cubic lattices. Consider the 3D hexagonal lattice which consists of two hexagonal layers with the wallpaper group $p6mm$ stacked on top of each other. The TRI momenta comprise two copies of the ones on the 2D hexagonal lattice, separated by a perpendicular translation. Accordingly, the phases can easily be obtained by considering  the $k_{z}=0$ plane (Table \ref{2dTable}), which contains the ${\Gamma}$, ${M}$ and ${K}$ points, and those of the other translated plane associated with the points ${A}$, ${L}$ and ${H}$, respectively (Table \ref{3dTable}).  Consequently, there are eight TRS protected phases resulting from the combinations of a TRS protected phase in one plane and a trivial or a valley configuration in the other plane. Additionally,
there are two valley phases which are configurations with one plane featuring a valley phase and the other a trivial configuration. Notice that a potential double valley phase with a valley phase in each of the planes is not protected by 3D crystal symmetry and is therefore trivial.
We again point out that the truly 3D valley phase is determined by a three-dimensional point group, i.e., the one whose action cannot be reduced to the 2D case.

Most experimentally observed TBIs are of the $\Gamma$ kind, such as $\text{Bi}_2\text{Se}_3$ \cite{xia2009} and $\text{Bi}_2\text{Te}_3$\cite{chen2009}, except for $\text{Bi}_x\text{ Sb}_{1-x}$ where the 3D pursuit started \cite{hsieh2008,hsieh2009} which is  $r{\bar 3}m-T-r{\bar 3}_L$. The theoretically predicted rocksalt actinides \cite{zhangfelser2012} are actually of the translationally-active class $fm{\bar 3}m-T-f3(4)$. $\text{Sn}\text{Te}$, as well as {\text Sn}-doped compounds $\text{Pb}\text{Te}$ and $\text{Pb}\text{Se}$   have the same space group but the phase recently observed in these compounds has no TRS protection \cite{ando2012,xu2012,story2012} and is a valley phase \cite{fu2011}. Let us inspect this phase in more detail. This phase turns out to be indexed as $fm{\bar 3}m-f3(4)$ (Table \ref{3dTable}). The system has mirror planes in the momentum space formed by the $\Gamma$ and any two of the $L$ points, which thereby relate the remaining two $L$ points by symmetry. As a result, a mirror-symmetric crystal cut along ${\bar \Gamma}-{\bar X}-{\bar\Gamma}$ line in the (001) surface features a pair of Dirac cones (a double Dirac cone) which is therefore also protected by the same symmetry\cite{fuNatComm2012}. Notice that in addition we predict valley phases at the $W$ and the $U$ points in the Brillouin zone protected by both the four-fold and the three-fold rotational symmetries, labelled by $fm{\bar 3}m-f43_W$ and $fm{\bar 3}m-f43_U$, respectively. The $W$-phase originates from six inequivalent symmetry-related $W$-points in the BZ where a band inversion gives rise to a valley phase. In addition, in the same phase, the $(001)$ surface features Dirac cones for the cut along  ${\bar \Gamma}-{\bar M}-{\bar\Gamma}$ and ${\bar \Gamma}-{\bar S}-{\bar\Gamma}$  lines, but not along ${\bar\Gamma}-{\bar X}-{\bar\Gamma}$ direction, as it is the case in the  $fm{\bar 3}m-f3(4)$ phase, see Fig.\ 2C. Therefore, the detection of the Dirac cones  in the ${\bar \Gamma}-{\bar M}-{\bar\Gamma}$ and ${\bar \Gamma}-{\bar S}-{\bar\Gamma}$ directions  in ARPES experiments would be a clear signature of this valley phase.

In conclusion, we provided the space group classification of the TBIs both in 2D and 3D. As a result, we found two additional broad classes of topological phases, besides the TRS protected $\Gamma$ states: the translationally-active phases, protected by both TRS and crystal symmetries, but susceptible to topological crystalline disorder, and the valley phases solely protected by the space group symmetry, and therefore susceptible to both elastic and topological crystalline disorder. Our complete classification scheme based on the full 2D and 3D space groups has as most important consequence that it demonstrates the potential existence of at least seventy distinct topological phases of insulating matter and we anticipate that this will be a valuable guide in the future exploration of this landscape.

\section*{Acknowledgements}
This work is supported by the Dutch Foundation for Fundamental Research on Matter (FOM).  V.\ J.\ acknowledges the support of the Netherlands Organization for Scientific Research (NWO).

\section*{Author Contributions}
All authors contributed extensively to the work presented in this paper.

\newpage
\clearpage
\section{Tables}
\begin{table}[h]
\footnotesize
\begin{tabular}{c|c|c|c|c}
\hline
\hline
{ Bravais Lattice (PG)} & { WpG}& ${\Gamma}_{i}$ &{$\delta_i$}&
Index (Phase)  \\
\hline
{Square}{ ($D_{4}$)} & $p4mm$ & (${\Gamma}$,
${M}$, ${X}$, ${Y}$) &(-1,1,1,1) &$T\text{-}p4mm$
($\Gamma$)\\ \cline{4-5}
&$p4gm$ &  &(1,-1,1,1) &$T\text{-}p4$ ($M$)\\ \cline{4-5}
 &$p4$ & &(1,1,-1,-1) &$p4$ ($X$-$Y$-valley)\\ \cline{4-5}
 \hline
 {Rectangular}{ ($D_{2}$)} & $p2mm$ & (${\Gamma}$,
${M}$, ${X}$, ${Y})$&(-1,1,1,1)
&$T\text{-}p2mm$ ($\Gamma)$\\ \cline{4-5}
& $p2mg$  & &(1,-1,1,1)&$T\text{-}p2m_{M}$ ($M$)\\ \cline{4-5}
&  $p2gg$ & &(1,1,-1,1) &$T\text{-}p2m_{X}$ ($X$)\\ \cline{4-5}
& $pm$, $pg$  & &(1,1,1,-1) &$T\text{-}p2m_{Y}$ ($Y$)\\ \cline{4-5}
\hline
{Rhombic}{ ($D_{2}$)} & $c2mm$ & $({\Gamma}$,
${M}_{0}$, ${M}_{-1}$, ${M}_{1})$&(-1,1,1,1) &$T\text{-}c2mm$ ($\Gamma$)\\ \cline{4-5}
&$cm$  & &(1,-1,1,1) &$T\text{-}c2m$ ($M_0$)\\ \cline{4-5}
& & &(1,1,-1,-1)&$c2m$ ($M$-valley)\\ \cline{4-5}
\hline
{Oblique}{ ($C_{2}$)} &$p2$&$({\Gamma}$,
${M}_{0}$, ${M}_{-1}$, ${M}_{1})$ &(-1,1,1,1)
&$T\text{-}p2$ ($\Gamma$)\\ \cline{4-5}
&$p1$  &&(1,-1,1,1)&$T\text{-}p2_{M_{0}}$ ($M_0$)\\ \cline{4-5}
& & &(1,1,-1,1) &$T\text{-}p2_{M_{\text{-}1}}$ ($M_{-1}$)\\ \cline{4-5}
& & &(1,1,1,-1)&$T\text{-}p2_{M_{1}}$ ($M_1$) \\ \cline{4-5}
\hline
{Hexagonal} & $p6mm$ & $({\Gamma}$,
${M}_{0}$, ${M}_{-1}$, ${M}_{1}$,
${K}_{-}$, ${K}_{+}$)&(-1,1,1,1,1,1) &$T\text{-}p6mm$ ($\Gamma$)\\ \cline{4-5}
 {\it(hexagonal -- $D_{6}$)}&$p6$ & &(1,-1,-1,-1,1,1) &$T\text{-}p6$ ($M$)\\ \cline{4-5}
& & &(1,1,1,1,-1,-1)&$p6$ ($K$-valley)\\ \cline{2-5}
Hexagonal& $p3m1$&$({\Gamma}$,
${M}_{0}$, ${M}_{-1}$, ${M}_{1}$)&(-1,1,1,1)&$T\text{-}p3m1$ ($\Gamma$)\\
 {\it(rhombohedral -- $D_{3}$)} & $p31m$, $p3$  & &&\\
\hline
\hline
\end{tabular}
\caption{Table of the topological phases in two dimensions. For each of the lattice structures, the corresponding point-group (PG) symmetry and the relevant wallpaper group (WpG), i.e. space group, are given. The corners of the square and rectangle are denoted by $M$, whereas in the triangular Bravais structure they are indicated by $K$. Additionally, the centers of the edges are denoted by $X$ and $Y$ in both the square and rectangular case and by $M$ in the other lattices\cite{dresselhausbook}. The resulting phases are characterized by the distribution of $\delta_i$ at the $\Gamma_i$ points consistent with the WpG symmetry. Phases cluster in Bravais lattices, with the hexagonal structure being the only exception. In this case the WpGs containing six-fold and three-fold rotational symmetries relate the high symmetry points in different ways. As a result, the Hamiltonian does not commute with the time-reversal operator at the $K$ points in the latter case. The obtained phases are ultimately protected by TRS  (whenever $\nu=1$), WpG symmetry, or both, and are accordingly indexed. The index (last column) describes the part of wallpaper group that leaves the subset $\Gamma_i$ having $\delta_{i}=-1$ invariant, while the additional label 'T' denotes TRS protection. In the column denoted ``Phase'' we introduce a convenient but imprecise shorthand notation.}
\label{2dTable}
\end{table}

\newpage
\clearpage
\begin{table}[h]
\footnotesize
\renewcommand{\arraystretch}{1.1023}
\begin{tabular}{c|c|c|c|c|c}
\hline
\hline
{ Bravais Lattice} & { PGS} & { SG}& ${\Gamma}_{i}$&
$\delta_{i}$&{ Index (Phase)}  \\
\hline
{Primitive Cubic} &$O_{h}$& $pm\bar{3}m$ &(${\Gamma}$, ${R}$,
${X}$, ${M}$)  &(-1,1,1,1)&$T\text{-}pm\bar{3}m$ ($\Gamma$)\\ \cline{5-6}
&& $pn\bar{3}n$   & &(1,-1,1,1) & $T\text{-}p3(4)_{R}$ ($R$)\\ \cline{5-6}
& &$pn\bar{3}m$ &  &(1,1,-1,1)&$T\text{-}p3(4)_{X}$ ($XYZ$)\\ \cline{5-6}
& &$pm\bar{3}n$ & &(1,1,1,-1) &$T\text{-}p3(4)_{M}$ ($MX'Y'$)\\
\hline
{Hexagonal}&$C_{6v}$& $p6mm$  & (${\Gamma}$, ${M}$, ${A}$, ${L}$, ${K}$, ${H}$)& (-1,1,1,1,1,1)&$T\text{-}p6mm$ ($\Gamma$) \\ \cline{5-6}
&&$p6cc$& &(1,-1,1,1,1,1)&$T\text{-}p6_{M}$ ($M$)\\ \cline{5-6}
&&$p6_{3}cm$&&(1,1,-1,1,1,1)&$T\text{-}p6_{A}$ ($A$)\\  \cline{5-6}
&&$p6_{3}mc$&&(1,1,1,-1,1,1)&$T\text{-}p6_{L}$ ($L$)\\  \cline{5-6}
&&&&(-1,1,1,1,1,-1)&$T\text{-}p6_{\Gamma H}$ ($\Gamma H$)\\ \cline{5-6}
&&&&(1,-1,1,1,1-1)&$T\text{-}p6_{HM}$ ($MH$)\\ \cline{5-6}
&&&&(1,1,-1,1,-1,1)&$T\text{-}p6_{KA}$ ($KA$)\\ \cline{5-6}
&&&&(1,1,1,-1,-1,1)&$T\text{-}p6_{LK}$ ($LK$)\\ \cline{5-6}
&&&&(1,1,1,1,-1,1)&$p6_{K}$ ($K$-valley)\\ \cline{5-6}
&&&&(1,1,1,1,1-1)&$p6_{H}$ ($H$-valley)\\
\hline
{Face Centered Cubic}&$O_{h}$& $fm\bar{3}m$  & (${\Gamma}$, ${X}$, ${L}$, ${U}$, ${W}$)& (-1,1,1,1,1)&$T\text{-}fm\bar{3}m$ ($\Gamma$) \\ \cline{5-6}
&& $fm\bar{3}c$  && (1,-1,1,1,1) &$T\text{-}f3(4)$ ($X$)\\ \cline{5-6}
&& $fd\bar{3}m$  && (1,1,-1,1,1)&$f3(4)$ ($L$-valley) \\ \cline{5-6}
&& $fd\bar{3}c$  & &(1,1,1,-1,1) &$f43_{U}$ ($U$-valley)\\ \cline{5-6}
& & & &(1,1,1,1,-1)&$f43_{W}$ ($W$-valley)\\ \hline
{Rhombohedral}&{$D_{3d}$}&$r\bar{3}m$&(${\Gamma}$, ${L}$, ${F}$, ${Z}$, ${P}$, ${K}$, ${B}$)&(-1,1,1,1,1,1,1)&$T\text{-}r${\={3}}$m$ ($\Gamma$)\\ \cline{5-6}
&&$r\bar{3}c$&&(1,-1,1,1,1,1,1)&$T\text{-}r${  \={3}}$_{L}$ ($L$)\\ \cline{5-6}
&&&&(1,1,-1,1,1,1,1)&$T\text{-}r${\={3}}$_{F}$ ($F$)\\ \cline{5-6}
&&&&(1,1,1,-1,1,1,1)&$T\text{-}r${\={3}}$_{Z}$ ($Z$)\\ \cline{5-6}
&&&&(1,1,1,1,-1,1,1)&$r${\={3}}$_{P}$ ($P$-valley)\\ \cline{5-6}
&&&&(1,1,1,1,1,-1,1)&$r${\={3}}$_{K}$ ($K$-valley)\\ \cline{5-6}
&&&&(1,1,1,1,1,1,-1)&$r${\text{\={3}}}$_{B}$ ($B$-valley)\\ \cline{5-6}
\hline
\hline
\end{tabular}
\caption{Topological phases anticipated in 3D for some specific point group (PG) symmetries.  Bravais lattices with same PG symmetries have different space groups (SG).  We point out that, in contrast to the 2D case, the phases do not cluster in Bravais lattice structures. For example, the fourfold rotational symmetry crucial for the  $f43_{U}$ and $f43_{W}$ phases is not contained in every space group associated with the face centered cubic lattice. As  $\delta_{i}$'s
attain the same value at the points ${\Gamma}_{i}$ related by lattice symmetry or a reciprocal lattice vector, only one representative is given from each set of such points.  We note that the rhombohedral $T\text{-}r${\={3}}$_{L}$ phase is observed in $\text{Bi}_{x}\text{Sb}_{1-x}$\cite{hsieh2008,hsieh2009}, while the $T\text{-}r${\={3}}$m$ phase is found in $\text{Bi}_{2}\text{Se}_{3}$ \cite{xia2009} and $\text{Bi}_{2}\text{Te}_{3}$\cite{chen2009}. Moreover, the $fm{\bar 3}m-f3(4)$ phase has recently been observed in $\text{Sn}\text{Te}$ \cite{ando2012}, as well as in {\text Sn}-doped compounds $\text{Pb}\text{Te}$ \cite{xu2012} and $\text{Pb}\text{Se}$ \cite{story2012}.}
\label{3dTable}
\end{table}

\newpage
\clearpage
\section{Figures }
\begin{center}
\begin{figure*}[h]
 \includegraphics[scale=0.8432]{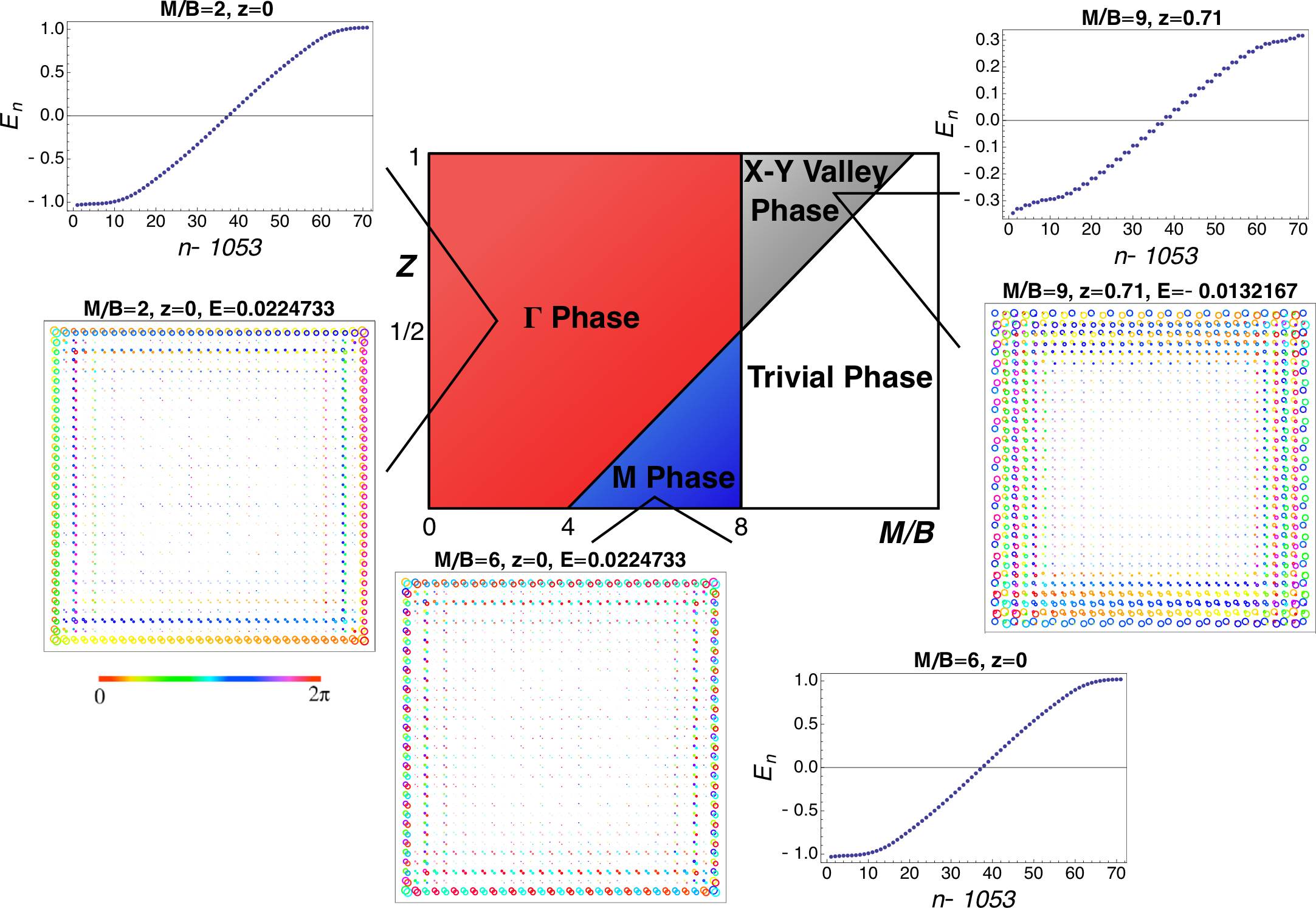}
\caption{Phase diagram of the extended $M-B$ tight-binding model. As function of the model parameters $M/B$ and $z=\tilde{B}/B$, where $B$ (${\tilde B}$) is the (next) nearest-neighbor hopping parameter and $M$ the difference in on-site energies, the different distributions of $\delta_i$ are obtained with the corresponding phases listed in Table \ref{2dTable}; consult SOM A for details.
Furthermore, the spectra of edge states per spin component are displayed for the non-trivial phases, demonstrating that the valley $X$-$Y$ phase exhibits a pair of Kramers pairs of metallic edge states. The real space localization of these edge states is also presented, where the radii of the circles represent the magnitude of the wave function and the colors indicate the phases as shown on the left. }
\label{figurephasesquare}
  \end{figure*}
  \end{center}
\newpage
\clearpage

\newpage
\begin{center}
\begin{figure*}[h]
\center
\includegraphics[scale=0.65]{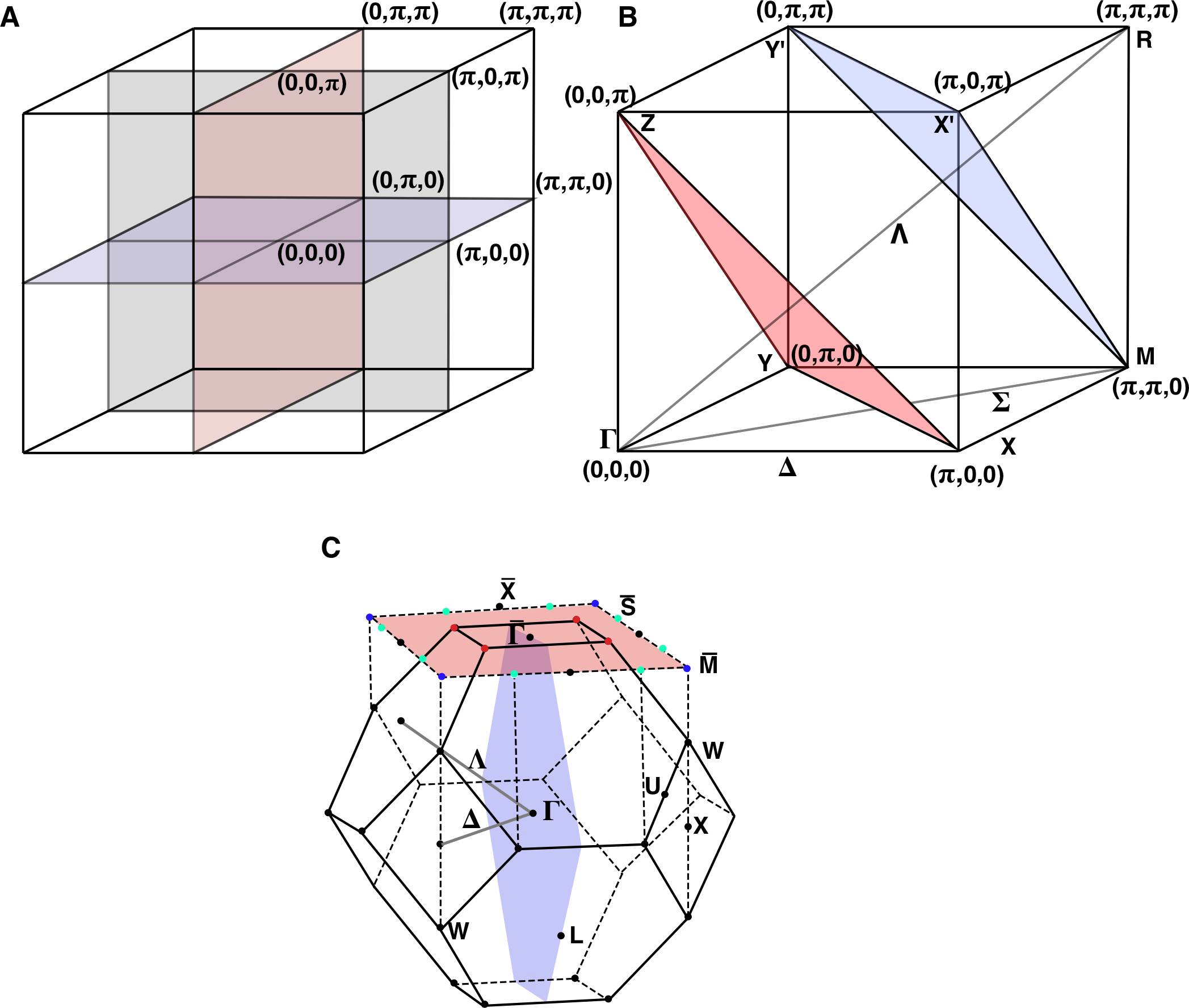}
	\caption{Illustration of the role of lattice symmetries in the classification of topological states. ({\bf A}) The eight TRI momenta in the Brillouin zone of the primitive cubic lattice. When only TRS is considered the sign of any quadruple of $\delta_{i}$'s within a plane connecting them can be changed, leaving their product the same. As a result one obtains, in addition to the 'strong' invariant, three weak invariants corresponding to the orthogonal planes. ({\bf B}) The constraints on the ${\delta}_i$'s arising from the lattice symmetries.  The high symmetry axes  $\Delta$, $\Lambda$ and $\Sigma$
represent axes of four-, three- and two-fold rotations, respectively; these transform the TRI points in the colored planes into each other, and thus constrain the corresponding $\delta_{i}$'s to be equal. ({\bf C}) The Brillouin zone of the face centered cubic lattice with high symmetry points and a mirror plane that projects onto the ${\bar\Gamma}-{\bar X}-{\bar\Gamma}$ line in the $(001)$ plane. The $W$-valley phase features Dirac cones along ${\bar\Gamma}-{\bar M}-{\bar\Gamma}$ and ${\bar\Gamma}-{\bar S}-{\bar\Gamma}$ lines, but \emph{not} along ${\bar\Gamma}-{\bar X}-{\bar\Gamma}$ lines. }
\label{fig:cubic}
\end{figure*}
\end{center}

\newpage

\title{\bf \Large Supplementary Information for ``The space group classification of topological band insulators"}\\
\author
{\large Robert-Jan Slager$^{1}$, Andrej Mesaros$^2$, Vladimir Juri\v ci\' c$^{1}$  and Jan Zaanen$^1$\\
\\
\normalsize{$^1$}{Instituut-Lorentz for Theoretical Physics, Universiteit Leiden, P.O. Box 9506, 2300 RA Leiden, The Netherlands}\\
\normalsize{$^2$}{Department of Physics, Boston College, Chestnut Hill, MA 02467, USA}\\
 }
\date{}
\maketitle
\section*{A. Model details}
We present details on the tight-binding models used to illustrate the phases in two dimensions. Specifically, we consider models of the form as introduced in Ref.\ [{\it S1}] which we generalize to the five Bravais structures in two dimensions. These  models assume nearest neighbor hopping on the specific lattice, which consists of spin degenerate $|s\rangle$ and $|p_{x}+ip_{y}\rangle$ type orbitals on every single site, and are therefore, in natural units $\hbar=c=e=1$, described by
\begin{equation}\label{Hamiltonian::TB}
\mathcal{H}_{\text{TB}}=\sum_{\mathbf{R}, \boldsymbol{\delta}} (\Psi^{\dagger}_{\mathbf{R}} T_{\boldsymbol{\delta}}\Psi_{\mathbf{R}+\boldsymbol{\delta}}+\text{H.c.}) +\sum_{\mathbf{R}}  \Psi^{\dagger}_{\mathbf{R}} \epsilon\Psi_{\mathbf{R}}.\tag{S1}
\end{equation}
Here  $\{\boldsymbol{\delta}\}$ denotes the vectors connecting the nearest neighbors on a particular  Bravais lattice and $\Psi^\top_{\mathbf{R}}=(s_{\uparrow}(\mathbf{R}),p_{\uparrow}(\mathbf{R}), s_{\downarrow}(\mathbf{R}), p_{\downarrow}(\mathbf{R}))$ annihilates the $s$ and $p$
type orbitals at site $\mathbf{R}$. We set the hopping matrices for the spin up and spin down electrons
\begin{equation*}
T^{*}_{\boldsymbol{\delta}, \downarrow \downarrow}=T_{\boldsymbol{\delta}, \uparrow \uparrow}=
\begin{pmatrix}
\Delta_{s}& t_{\boldsymbol{\delta}}/2\\
t'_{\boldsymbol{\delta}}/2&\Delta_{p}
\end{pmatrix},
\end{equation*}
with $t_{\boldsymbol{\delta}}=-iA_{\boldsymbol{\delta}}\exp({i\varphi_{\boldsymbol{\delta}}})$ and  $t'_{\boldsymbol{\delta}}=-iA_{\boldsymbol{\delta}}\exp({-i\varphi_{\boldsymbol{\delta}}})$, in terms of the polar angle $\varphi_{\boldsymbol{\delta}}$ of the vector $\boldsymbol{\delta}$,  $\Delta_{s/p,{\boldsymbol\delta}}=D\pm B_{\boldsymbol{\delta}}$,
and on-site energies $\epsilon=[(C-4D)\tau_{0}+(M-4B)\tau_{3}]\otimes\sigma_{0}$, where the Pauli matrices ${\boldsymbol \tau}$ and ${\boldsymbol\sigma}$ act in the orbital and spin space, respectively. Here, ${\tau}_0$ and ${\sigma}_0$ are the $2\times2$ unity matrices.

By performing a Fourier transform, the Hamiltonian \eqref{Hamiltonian::TB} assumes  the form
\begin{equation}\label{Hamiltonian::TriRec}
\mathcal{H}_{\text{TB}}=\sum_{{\bf k}}\Psi_{\bf k}^\dagger
\begin{pmatrix}
H(\mathbf{k})&0\\
0&H^{*}(-\mathbf{k})
\end{pmatrix}\Psi_{\bf k}\tag{S2}
\end{equation}
where
\begin{equation*}
H(\mathbf{k})={\boldsymbol\tau}\cdot{\bf d}({\bf k})
\end{equation*}
and the vector ${\bf d}({\bf k})$ obviously depends on the form of the lattice. The upper and the lower block of this Hamiltonian are related by the time-reversal symmetry represented by an antiunitary operator ${\cal T}=\tau_0\otimes i\sigma_2 K$, with $K$ as the complex conjugation. The form of the tight-binding Hamiltonian in momentum space is quite generic in a sense that it is
valid also when longer-range hopping terms, such as next-nearest neighbor ones, are included in the Hamiltonian (\ref{Hamiltonian::TB}), as we will see later.
As a result, the spectrum of the above Hamiltonian takes the simple form in terms of the vector ${\bf d}$, $E(\mathbf{k})=\pm\sqrt{{\bf d}\cdot{\bf d}}\equiv\pm|{\bf d}|$. The closing of the bandgap occurs when  the magnitude of the vector ${\bf d}({\bf k})$ vanishes, i.e., when ${\bf d}({\bf k})=0$, which is the case {\it only} at the special high-symmetry points in the Brillouin zone (BZ) and for special values of the parameters in the Hamiltonian, as we will see in the specific examples.

Taking up the square lattice and neglecting the energy shifts, we obtain the familiar three component ${\bf d}(\mathbf{k})$ vector with $d_{1,2}(\mathbf{k})=\pm\sin(k_{x,y})$ and $d_{3}=M-2B(2-\cos(k_{x})-\cos(k_{y}))$.  The corresponding spectrum, doubly degenerate in the spin space, is gapped except for the values  $M/B= 0, 4$ or $8$. When $0<M/B<4$ the band gap opens at the $\Gamma$ point in the BZ and the system enters a topological phase, denoted as the $\Gamma$ phase. For $4<M/B<8$ the system is still topologically non-trivial ($M$ phase), but with the band gap opening at $\mathbf{k}=(\pi,0)$  ($X$ point) and $\mathbf{k}=(0,\pi)$ ($Y$ point). In contrast, the other values of $M/B$ give rise to a topologically trivial phase. Interestingly, the resulting nontrivial phases in two dimensions can (if needed by adiabatic continuity) be visualized by the position of the Skyrmion of the band structure field $\hat{\mathbf{d}}(\mathbf{k})\equiv\mathbf{d}(\mathbf{k})/|\mathbf{d}(\mathbf{k})|$ with the associated density $s (\mathbf{k})=\hat{\mathbf{d}}(\mathbf{k})\cdot [\partial_{k_{x}}\hat{\mathbf{d}}(\mathbf{k}) \times \partial_{k_{y}}\hat{\mathbf{d}}(\mathbf{k})]$.
Moreover, these phases can be probed by dislocations, and only the $M$-phase results in zero-modes bound to the dislocations, whereas a $\pi$ flux binds zero-energy modes in both non-trivial states [{\it S2}].

The connection with the characterization presented in the main text is achieved by relating the Pfaffian to the band-structure vector ${\bf d}(\bf k)$ through $\delta_{i}=\text{sign}[d_{3}(\mathbf{\Gamma}_{i})]$, which can be readily shown, as well as by noting that
$\Gamma_{i}$ comprise the points at which the gap closes. As a result the $\Gamma$ and $M$ phase are indeed characterized by an opposite sign of the Pfaffian at those points.
Furthermore, the $C_4$ rotation connects the $\delta_{i}$'s and thus dictates that the values at the $X$ and $Y$ points have  to be the same. As a result, the $\delta_{i}$'s can assume a configuration for which $\delta_X=\delta_Y\neq\delta_\Gamma=\delta_M$, which we call the $X-Y$ valley phase. This phase can easily be incorporated in the model \eqref{Hamiltonian::TB} by introducing a  next nearest neighbor term with the parameters defined exactly as in the nearest neighbor case. Accordingly, we obtain \begin{equation}\label{eq::d(k)extendedsquare}
\mathbf{d}(\mathbf{k})=
\begin{pmatrix}
\sin(k_{x})+\cos (k_x)\sin( k_y)\\
-\sin(k_{y}) -\sin (k_x )\cos (k_y)\\
M-2B[2-\cos (k_{x})- \cos (k_{y})]-4\tilde{B}[1-\cos (k_x)\cos( k_y)]
\end{pmatrix}.\tag{S3}
\end{equation}
The gap at the TRI momenta and hence the configuration of $\delta_{i}$ can be tuned as function of the parameters  $\tilde{B}/B$ and $M/B$  resulting in the phase diagram shown in Fig. 1 in the main text.
Importantly, we find that the resulting  valley phase features two pairs of edge states (Fig. 1 in the main text). In this case there are two single Skyrmions located at the X and Y points and integration of the Skyrmion density yields a winding number two per spin. Moreover, this phase can also be probed in the exact same manner as the $\Gamma$ and $M$ phases (see Supplementary Information, Section B).

As the next step, the rectangular model is obtained by making the magnitude of the hopping parameters anisotropic, which effectively reduces the fourfold rotational symmetry to the twofold, see also Section B in Supplementary Information, Eq.\ \eqref{eq::hamiltonianrectangle}.
Consequently, the oppositely valued $\delta_{i}$'s can be located at any of the TRI momenta $\Gamma,M,X,Y$, and the corresponding phases can be distinguished by insertion of dislocations (see Supplementary Information, Section B). We find that by increasing the ratio $r$ of hopping magnitudes in the ${y}$ and ${x}$ directions the valley phase is lost: the system either enters a trivial state, a translationally-active $X$ phase or a phase with $C=2$ per spin not protected by either TRS or lattice symmetry. Namely, only the former two phases are realized when the ratio $r$ is large enough. The two edge modes per spin in the $X$-$Y$ valley phase are gapped out upon an infinitesimal breaking of $C_4$ symmetry, i.e. by $r>1$.
The phase diagram of the model for $r=2$ is shown in Fig.\ S5.
 \begin{figure*}[t!]
\centering
\includegraphics[clip,width=0.96\linewidth]{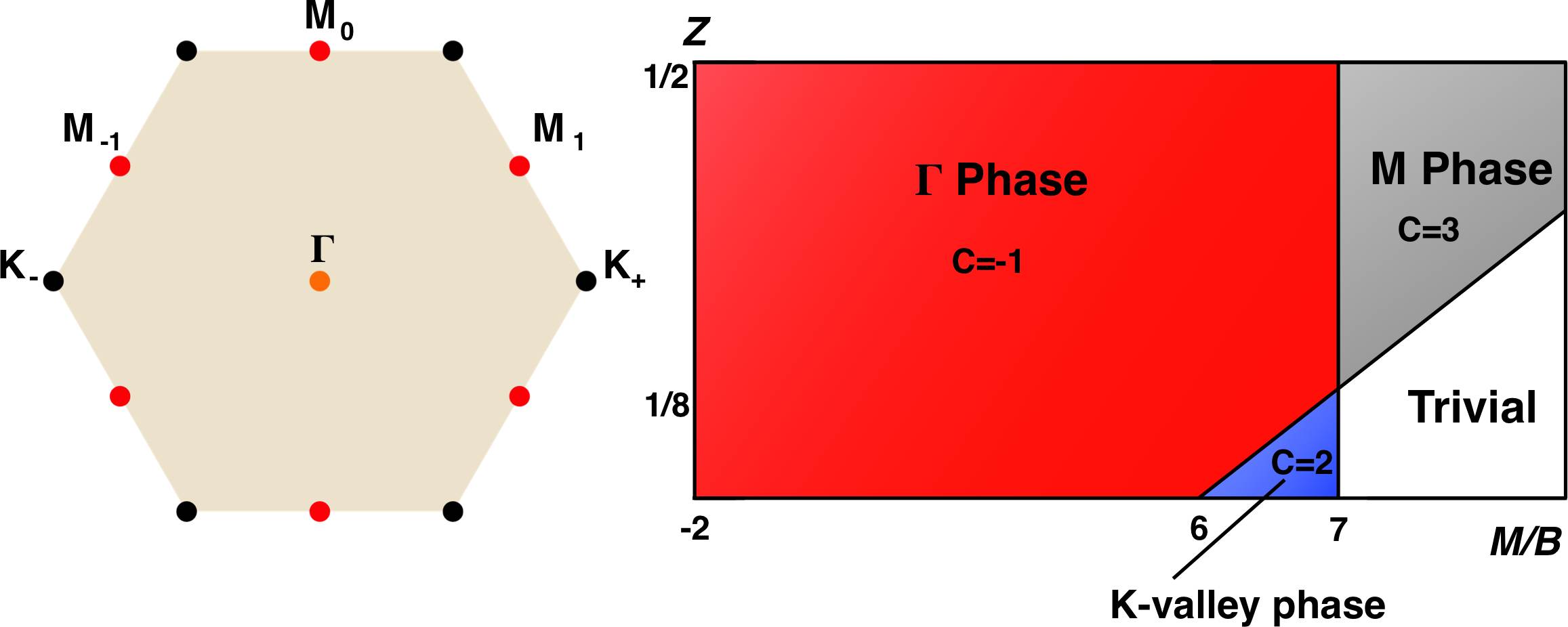}
\caption*{\label{figureS21} Figure S1: The Brillouin zone of the primitive hexagonal (triangular) lattice and the phase diagram of the $M$-$B$ tight-binding model on the same lattice. High-symmetry points in the Brillouin zone at which the Hamiltonian commutes with the time-reversal operator are indicated. Additionally, the Chern number per spin block ($C=C_\uparrow=-C_\downarrow$) is shown for each phase. }
\end{figure*}

In case of a primitive hexagonal (triangular) lattice with the wallpaper group $p6mm$ and with the Brillouin zone shown in Fig.\ S1, the nearest neighbor hopping part of the Hamiltonian is described by
\begin{equation}\label{Eq::Triangular}
{\bf d}(\mathbf{k})=
\begin{pmatrix}
\sin k_x+2\sin(\frac{k_x}{2})\cos(\pi/3)\cos(\frac{\sqrt{3}}{2}k_y)\\
-2\cos(\frac{k_x}{2})\sin(\pi/3)\sin(\frac{\sqrt{3}}{2}k_y)\\
M-2B[2-\cos k_x-2\cos(\frac{k_x}{2})\cos(\frac{\sqrt{3}}{2}k_y)].\\
\end{pmatrix}\tag{S4}
\end{equation}
As for the other Bravais lattices, the sign of the Pfaffian at the $\Gamma_i$ points is $\delta_{i}=\text{sign}[d_{3}({\Gamma}_{i})]$
and
we observe that due to the very presence of the lattice symmetry ($C_6$ in particular), $\delta_{i}$ is also well defined at the two inequivalent corners $K_{+}$ and $K_{-}$ of the Brillouin zone. Additionally, the symmetry relates the centers of the edges $M$ and we thus expect the phases as indicated in the Table 1 in the main text. The resulting Hamiltonian corresponding to the vector ${\bf d}$ in \eqref{Eq::Triangular} indeed exhibits the $\Gamma$ and the $K$ valley phase for $-2<M/B<6$ and $6<M/B<7$, respectively. The $M$ phase can also be captured by including the next nearest neighbor hopping term
\begin{equation}
\tilde{\mathbf{d}}(\mathbf{k})=
\begin{pmatrix}
\tilde{A}\sin(\frac{3}{2}k_{x})\cos(\pi/6)\cos(\frac{\sqrt{3}}{2}k_{y})\\
-\tilde{A}[\sin(\sqrt{3}k_{y})-2\cos(\frac{3}{2}k_{x})\sin(\pi/6)\sin(\frac{\sqrt{3}}{2}k_{y})]\\
-2\tilde{B}[3-\cos(\sqrt{3}k_{y})-2\cos(\frac{3}{2}k_{x})\cos(\frac{\sqrt{3}}{2}k_{y})]\tag{S5}
\end{pmatrix}.
\end{equation}
 For strong enough next-nearest neighbor  hopping, $z=\tilde{B}/B>\frac{1}{8}$,  the gap closes at the $K$ points before closing at the $M$ points, for increasing $M/B$, and as a result the system exhibits a $\Gamma$ and $M$ phase, see Figs. S1 and S2.
\begin{figure*}[t!]
\includegraphics[scale=0.163]{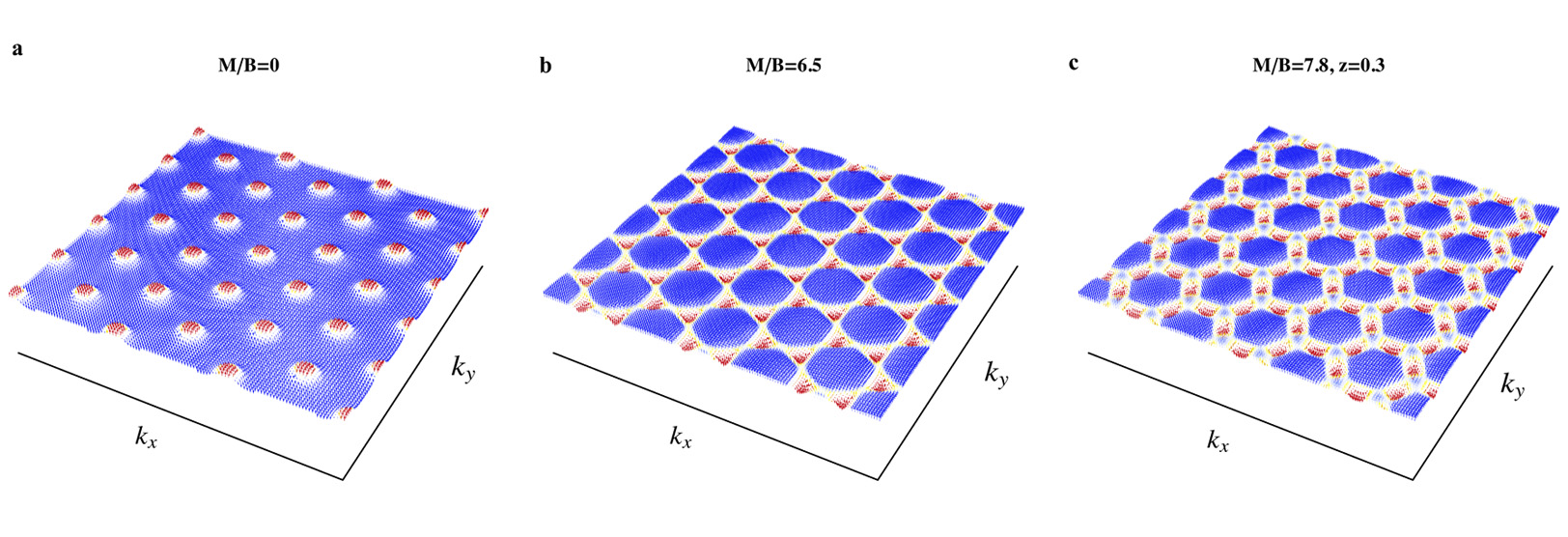}
\caption*{\label{figureS2} Figure S2: The skyrmion lattices in the extended Brillouin zone for the three topological phases on the primitive hexagonal (triangular) lattice. The locus of the Skyrmions is colored in red. (a) The Skyrmion lattice for the $\Gamma$ phases within the model \eqref{Hamiltonian::TB}. (b) $K$ valley phase within the same model. (c) When next nearest neighbor hopping, $z=\tilde{B}/B$, is taken into account, the system can enter the $M$ phase, which has the Skyrmions positioned at the $M$ points. }
\end{figure*}
We note that we need six inequivalent high symmetry points, marked in Fig. S1, to capture the three observed phases within the rule presented in the main text. In contrast, the symmetry relating $K_{+}$ and $K_{-}$ can readily be broken by considering a non-primitive honeycomb lattice structure, realized in graphene, which then makes $K_\pm$ {\emph{ related}} by TRS.  The Hamiltonian does not commute with the time-reversal operator at these points, and therefore, these two points no longer belong to the set ${\Gamma}_{i}$. As the $M$ points are still related by symmetry, we expect only a $\Gamma$ phase in this case. A prominent example is the Kane-Mele model [{\it S3}] with the configuration of ${\delta}_i$'s explicitly calculated in Ref.\ [{\it S4}] using a rhombic unit cell. Drawing this cell in the extended BZ we observe that this configuration indeed corresponds with a $\Gamma$ phase in the usual Wigner-Seitz cell. Additionally, the result can also be understood from the Skyrmion picture. Due to inversion-asymmetry the gap closes at $K_\pm$ points instead of closing at the TRI momenta [{\it S5, S6}]. The spin sub-blocks  exhibit  merons at $K_{+}$ and $K_{-}$ points which can be explicitly shown by rewriting the Pontryagin index as in Ref.\ [{\it S7}].

Finally, we consider the oblique and rhombic Bravais structures distinguished by the presence of mirror symmetry in the latter case. This symmetry is ultimately responsible for the existence of the valley phase on the rhombic lattice. Considering the rhombic lattice, it can be readily shown that
\begin{equation}
\small
\mathbf{d}(\mathbf{k})=
\begin{pmatrix}
A_{1} \cos(\phi_1)\sin(\tilde{k}_{x})\cos(\tilde{k}_{y})+A_{2}\sin(r_1k_{x})\\
-A_{1} \sin(\phi_1)\sin(\tilde{k}_{x})\cos(\tilde{k}_{y})-A_{3}\sin(r_2k_{y})\\
M-4B[1-\cos(\tilde{k}_{x})\cos(\tilde{k}_{y})]+2\tilde{B}[\tau_{1}^{-1}+\tau_{2}^{-1}-\tau_{1}^{-1}\cos(r_{1}k_{x})-\tau_{2}^{-1}\cos(r_{2}k_{y})]
\end{pmatrix},\tag{S6}
\end{equation}\label{eq:rhombicHam}
where $\tilde{k}_{x}=e_1\cos(\phi_1)k_{x}$, $\tilde{k}_{y}=e_1\sin(\phi_1)k_{y}$, with $\phi_1$ as the polar angle of the vector ${\bf e}_1$, and $e_1=|{\bf e}_1|$; see Fig. S3. Furthermore, $A_i$ is the hopping between the $s$ and the $p$ orbitals along the vector ${\bf e}_i$, while $B$ is the hopping between the two $s$ orbitals along the vector ${\bf e}_1$, and the hopping between these orbitals along the vectors ${\bf e}_{2,3}$ is $B_{2,3}={\tilde B}\tau_{2,3}^{-1}$.  Due to the mirror symmetry, the $M_{-1}$ and $M_{1}$ points are connected and as a result the gap closes at these points for the same value of $M/B$. Consequently, tuning the gaps at the TRI momenta by varying the parameters of the model,  the configurations of $\delta_i$'s and the corresponding phases shown in Table 1 in the main text are easily obtained.
\begin{figure*}[t!]
\centering
\includegraphics[clip,width=0.52\linewidth]{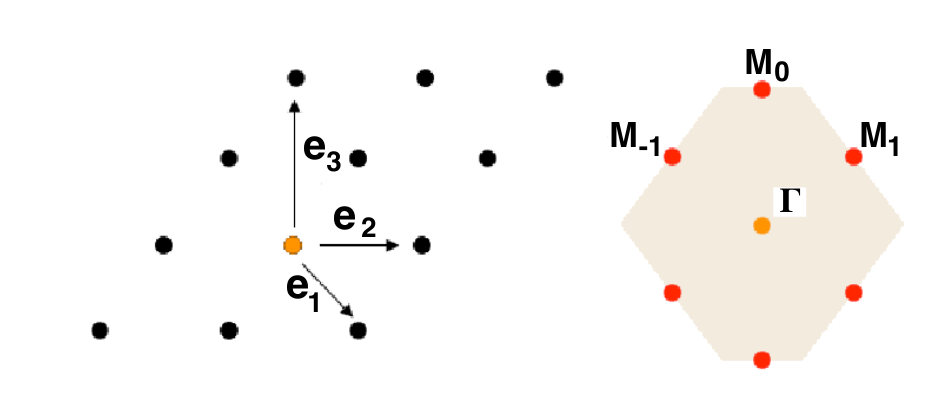}
\caption*{\label{figureS3} Figure S3: The left panel shows the real space rhombic lattice with the vectors connecting the neighbors. The right panel shows the corresponding Brillouin zone with the TRI momenta for  $e_2=|{\bf e}_2|=3$ and $e_3=|{\bf e}_3|=4$ in arbitrary units. Crucially, the mirror symmetry connects the $M_{-1}$ and $M_{1}$ points.}
\end{figure*}
The oblique lattice can be then treated similarly. Denoting the different connections between neighbors as indicated in Fig. S4,  we obtain
\begin{equation}
\mathbf{d}(\mathbf{k})=
\footnotesize
\begin{pmatrix}
A_{1}\cos(\phi_1)\sin(\mathbf{e}_{1}\cdot\mathbf{k})+A_{2}\cos(\phi_{2})\sin(\mathbf{e}_{2}\cdot\mathbf{k})+A_{3}\cos(\phi_{3})\sin(\mathbf{e}_{3}\cdot\mathbf{k})+A_{4}\cos(\phi_{4})\sin(\mathbf{e}_{4}\cdot\mathbf{k})\\
-A_{1}\sin(\phi_1)\sin(\mathbf{e}_{1}\cdot\mathbf{k})-A_{2}\sin(\phi_{2})\sin(\mathbf{e}_{2}\cdot\mathbf{k})-A_{4}\sin(\phi_{4})\sin(\mathbf{e}_{4}\cdot\mathbf{k})\\
M-2B[2-\tau_{1}^{-1}\cos(\mathbf{e}_{1}\cdot\mathbf{k})-\tau_{2}^{-1}\cos(\mathbf{e}_{2}\cdot\mathbf{k})-\tau_{3}^{-1}\cos(\mathbf{e}_{3}\cdot\mathbf{k})-\tau_{4}^{-1}\cos(\mathbf{e}_{4}\cdot\mathbf{k})]
\end{pmatrix}.\tag{S7}
\end{equation}
where angles $\phi_i$, $i=1,...,4$, are the polar angles of  vectors ${\bf e}_i$. The $s-p$ hopping along the vector ${\bf e}_i$ is $A_i$, and the $s-s$ hopping along the same vector is $B_i=B \tau_i^{-1}$; see also Eq. (S1).
Again, by varying the model parameters (consequently also the shape of the BZ), it is straightforward to realize the different phases which are also visualized by the presence of a Skyrmion at either of the indicated points.
More interestingly, the analysis confirms yet again the significance of the symmetry group of the underlying crystal. Essentially, the only difference between the oblique and rhombic systems is the presence of the mirror symmetry. Therefore, the same $(1,1,-1,-1)$ configuration of $\delta_{i}$'s results in a valley phase with the Chern number equal two per spin block \emph{only} when the mirror symmetry is present, i.e., only in the case of the rhombic lattice.

\begin{figure*}[t!]
\centering
\includegraphics[clip,width=0.52\linewidth]{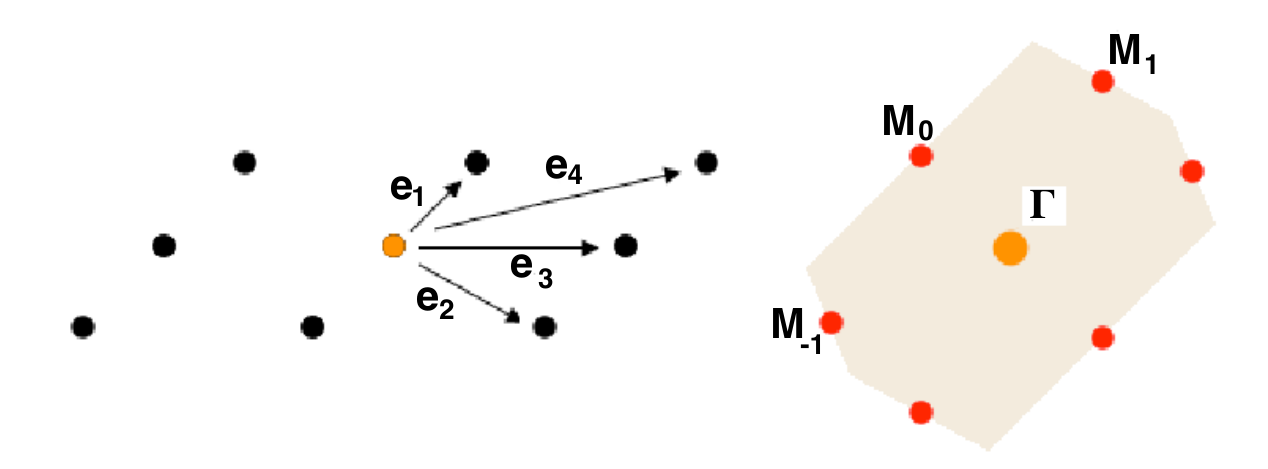}
\caption*{\label{figureS4} Figure S4: The real space oblique lattice and its corresponding Brillouin zone. In contrast to the rhombic case, the TRI momenta are not related by a lattice symmetry. In this example, vectors ${\bf e}_1$ and ${\bf e}_2$ connect nearest-neighboring sites, while  ${\bf e}_3$ and ${\bf e}_4$ connect next-nearest neighboring sites. }
\end{figure*}

\newpage
\section*{B. Probing the topologically non-trivial phases}

As already shown by the present authors within the $M$-$B$ tight-binding model on the square lattice, a $\pi$-flux  probes $Z_2$ topological order through binding of the zero modes, while the dislocations act as universal probes of the translationally-active topological states [{\it S2}]. Here we show that indeed the same holds for the valley phase on the square lattice, realized within an extended $M$-$B$ tight-binding model which includes next-nearest neighbor hopping. Moreover, we demonstrate by considering the same model on the rectangular lattice that the directionality of the dislocation can serve to distinguish translationally-active topological phases at different high symmetry points in the BZ.

The $M$-$B$ tight-binding  model \eqref{Hamiltonian::TB} on a rectangular lattice assumes the form (\ref{Hamiltonian::TriRec}) in terms of the vector
\begin{equation}\label{eq::hamiltonianrectangle}
\mathbf{d}(\mathbf{k})=
\begin{pmatrix}
\sin(k_{x})+\frac{\tilde{A}}{\sqrt{1+\tau}}\cos(\varphi_{r}) [\sin(k_{+})-\sin(k_{-})]\\
-1/\tau\sin(rk_{y})-\frac{\tilde{A}}{\sqrt{1+\tau}}\sin(\varphi_{r}) [\sin(k_{+})+\sin(k_{-})]\\
M-2B[2-\cos(k_{x})-1/r \cos(rk_{y})]-2\tilde{B}[2-\cos(k_{+})-\cos(k_{-})]
\end{pmatrix},
\tag{S8}
\end{equation}
where $r$ denotes the ratio of the lattice spacing in the ${y}$ and the ${x}$-direction,  $\varphi_{r}=\arctan(r)$, $\tau= A_{x}/A_{y}=B_{x}/B_{y}$, and $k_\pm=\pm k_x+r k_y$. In the remainder we conveniently fix $r=\tau=2$ and $\tilde{A}=1$. The resulting phase diagram is presented in Fig. S5.

\begin{figure*}[b!]
\centering
\includegraphics[clip,width=0.69\linewidth]{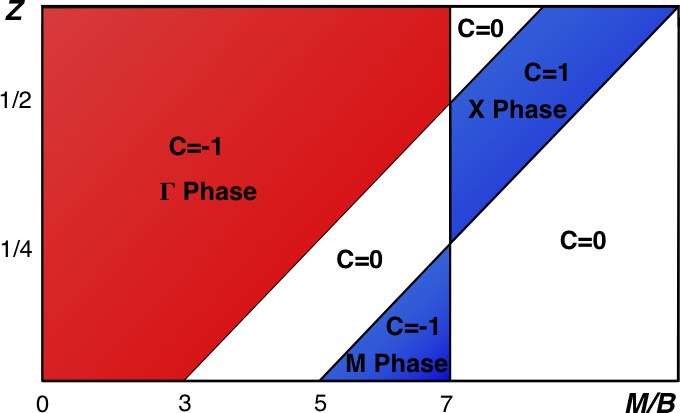}
\caption*{\label{figureS5} Figure S5: Phase diagram of the extended $M-B$ model on the rectangular lattice in Eq.\ \eqref{eq::hamiltonianrectangle} for $r=\tau=2$ and $\tilde{A}=1$. The nontrivial phases include the $\Gamma$, $M$ and a $X$ phases. Additionally, the Chern character $C$ per spin, $C=C_\uparrow=-C_\downarrow$ is indicated.}
\end{figure*}

\begin{figure*}[t!]
\centering
\includegraphics[clip,width=0.73\linewidth]{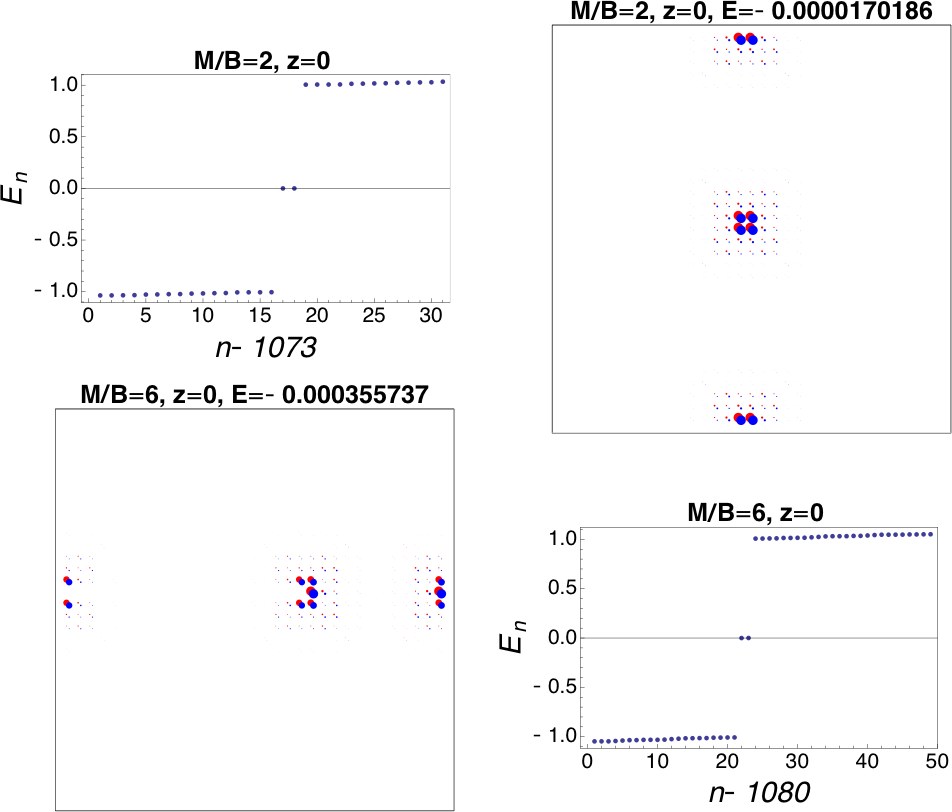}
\caption*{\label{figureS6} Figure S6: Effect of introducing a $\pi$ flux or dislocations in the $\Gamma$ and $M$ phase of the square lattice model. The top row shows two zero-modes bound to $\pi$-anti-$\pi$ flux pair (periodic boundary conditions) appearing in the spectrum in case of the $\Gamma$ phase and the real space localization of these modes. For convenience we only consider one spin component. The bottom row shows the effect of a dislocation-anti-dislocation pair (periodic boundary conditions) with Burgers vector $\mathbf{b}=\mathbf{e}_{y}$ in the $M$ phase. The lattices contain 33 $\times$ 33 unit cells.}
\end{figure*}

Considering first only nearest neighbor hopping, the non-trivial phases on both the square and rectangular lattices include
the $\Gamma$ and $M$ phases. Accordingly, we find that the $M$ phase gives rise to a dislocation zero-mode, whereas a $\pi$ flux results in a zero-mode in both the $\Gamma$ and the $M$ phase (Fig. S6). The dislocation zero mode hybridizes with the low-energy edge states in a finite system. Separating these states by introducing weak disorder via a random chemical potential, we confirm that the edge state indeed is at a finite momentum $k=\pi$.

\begin{figure*}[t!]
\centering
\includegraphics[clip,width=0.63\linewidth]{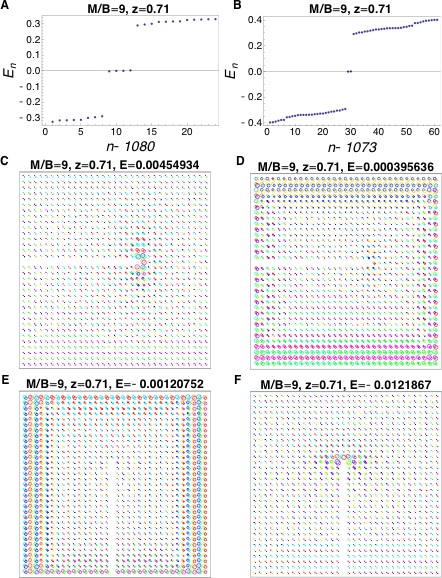}
\caption*{\label{figureS7} Figure S7: Probing the $X-Y$ valley phase, by considering one spin component of each Kramers pair. (a) Spectrum in presence of a $\pi$-flux and anti-$\pi$-flux in a periodic system. (b) Spectrum in case of a dislocation and anti-dislocation in a periodic system. (c) Real space localization of zero-mode resulting from a single dislocation with Burgers vector $\mathbf{b}=\mathbf{e}_{y}$. (d) The phase of the edge state hybridized with the dislocation zero mode displayed in (c), shows that it is a mode from the Y point, $\mathbf{k}=(0,\pi)$.
Similarly, the dislocation zero-mode displayed in (f) hybridizes with edge state shown in (e), the phases of which indicate that this zero-energy state corresponds with the X point, $\mathbf{k}=(\pi,0)$. Color-code is as in Fig. \ref{figurephasesquare}. The lattices contain 33 $\times$ 33 unit cells.}
\end{figure*}

The next nearest neighbor hopping, as already shown, realizes the valley phase on the square lattice (Fig. 1 in the main text). As anticipated, introduction of a magnetic $\pi$ flux results in a pair of Kramers pairs of zero modes in the system. More interestingly, the bulk-boundary correspondence and the directional character of a dislocation result in the possibility of discriminating between the two valley points. Namely, a dislocation with Burgers vector $\mathbf{b}$  along the $\hat{x}$ or $\hat{y}$ direction binds only a single pair of zero-energy modes. Analysis of the edge state hybridized with the dislocation zero-mode then confirms that the dislocation indeed ``sees'' the valley at the momentum parallel to the direction of the Burgers vector (Fig. S7).

When $r$ is increased (breaking square symmetry to rectangle), the Chern number per spin block may reduce to zero and insertion of a $\pi$-flux or a dislocation no longer results in a zero-mode in the system, confirming the intimate relation between the point group lattice symmetry and the valley phase. As the parameter $r$ is increased, another possibility is a topological phase transition to the $X$ phase which does not respect the fourfold symmetry and thus distinguishes the square from the rectangular lattice. This is reflected in the response of the system upon the insertion of dislocations. In addition to the usual zero-energy modes resulting from a $\pi$-flux vortex, the spectrum shows a pair of zero-modes resulting from dislocations only if the Burgers vector $\mathbf{b}$ has a component in the ${x}$ direction (Fig. S8).

Let us conclude by addressing the robustness of the zero-modes in the topological non-trivial regimes, which is related to the stability of the corresponding phase. To this end, we multiply all the model parameters  by Gaussian variables of width $w=10\%$ and also introduce Gaussian random chemical disorder, while preserving TRS.   Moreover, we couple the spin-reversed sub-blocks  by adding a nearest neighbor Rashba spin-orbit coupling
\begin{equation}
H_R=i\frac{R_{0}}{2}\sum_{\mathbf{R}, \boldsymbol{\delta}} \Psi^{\dagger}_{\mathbf{R}}[(\tau_{0}+\tau_{3})\otimes(\boldsymbol{\sigma}\times\boldsymbol{\delta})\cdot{\bf e}_z]\Psi_{\mathbf{R}+\boldsymbol{\delta}}
\tag{S9}
\end{equation}
to the Hamiltonian \eqref{Hamiltonian::TB}, which  breaks the reflection symmetry $z\rightarrow -z$ about the plane.
The results are displayed in Fig. S9.
\begin{figure*}[t!]
\center
\includegraphics[clip,width=0.59\linewidth]{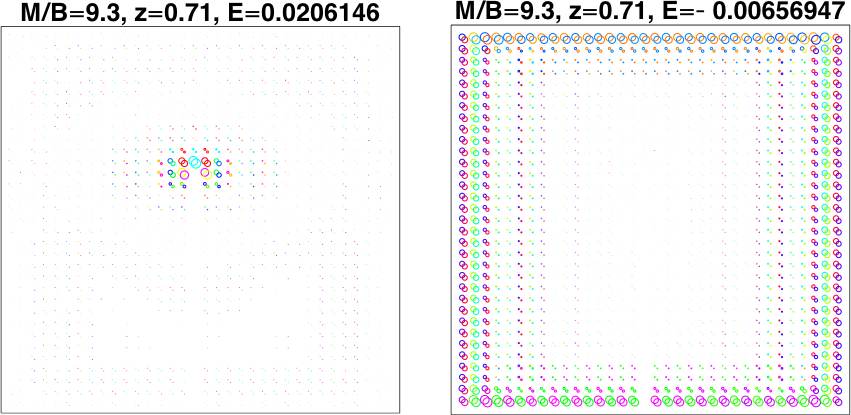}
\caption*{\label{figureS8} Figure S8: Illustration of the zero-mode bound to a dislocation appearing in the $X$ phase decoupled from the edge states using linear combinations (left panel). The right panel shows the edge state, the phases of which indicate that these edge states indeed correspond with the $X$ point (color-code as in Fig. \ref{figurephasesquare}). The lattice contains 30 $\times$ 33 unit cells. }
\end{figure*}
We find that finite Rashba spin orbit coupling, but not large enough to close the topological gap, preserves the real space localization of the zero-modes. More importantly, we observe that the stability of the modes  in the valley phase is significantly smaller than in the $Z_2$ non-trivial phases.  We note that the latter phase has a smaller gap and hence would already require a weaker Rashba coupling to close the gap. However,  one would expect the valley phase to be less stable independently of the gap size, as only the lattice symmetry accounts for its protection. To check this assertion, we also considered the valley phase in presence of a stronger next nearest neighbor hopping, which leads to an increase of the gap size. This analysis then confirms that the zero-modes in the valley phase are substantially less robust than those in the TRS protected phases.

\newpage

\begin{figure*}[t!]
\centering
\includegraphics[clip,width=0.93\linewidth]{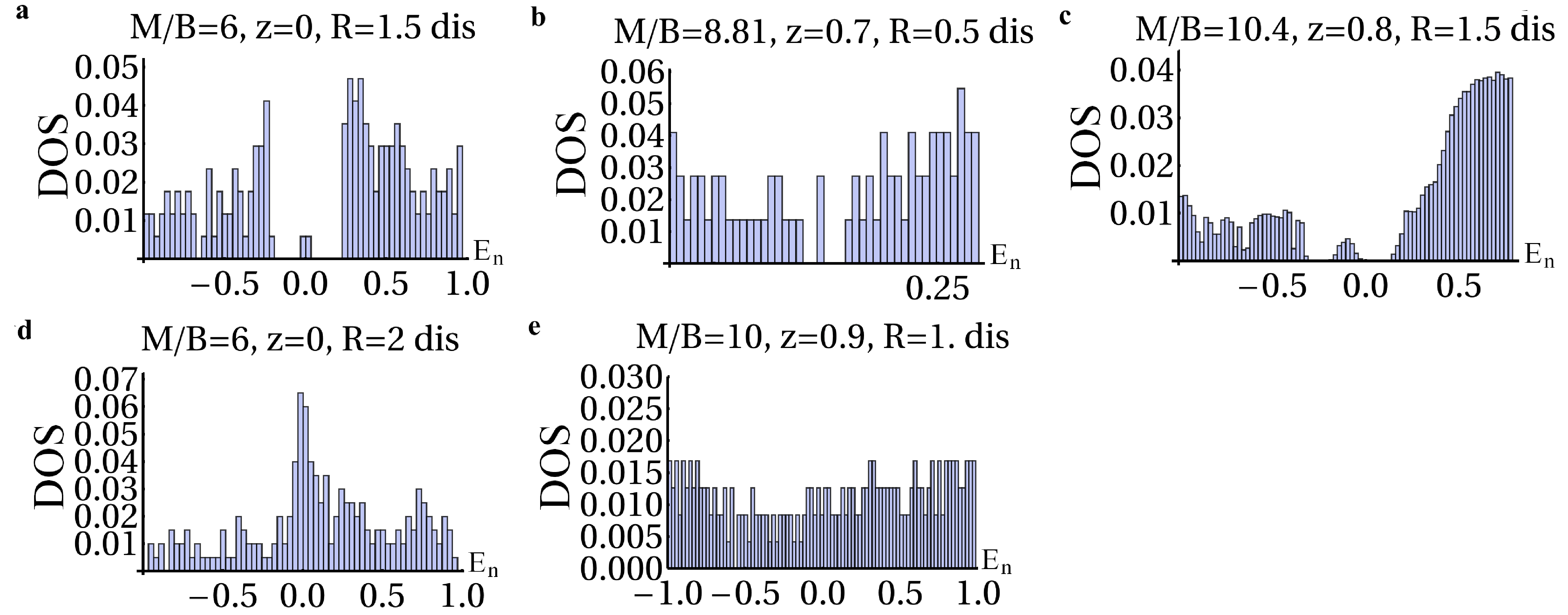}
\caption*{\label{figureS9} Figure S9: The averaged (over 130 disorder realizations) density of states (DOS) for various values of $M/B$, $z=\tilde{B}/B$ and $R=R_{0}/B$. Panels (a)-(b) and (d)-(e) correspond to the square lattice model, whereas panel (c) corresponds to the rectangular case. The midgap modes resulting from dislocations are present even for strong Rashba coupling and disorder, as demonstrated by (a) $R=1.5$  in the $M$ phase, (b) $R=0.5$ in the $X-Y$ valley phase and  (c) $R=1.5$ in the $X$-phase. Sufficiently strong Rashba coupling closes the topological gap, as shown for (d) $R=2$ in the $M$ phase (a similar result is obtained for the $X$-phase) and (e) $R=1$ in the $X-Y$ valley phase. Even for comparable values of the bulk gap, the valley phase is substantially less robust to the Rashba coupling than the TRS protected $M$ and $X$ phases.}
\end{figure*}

\section*{Supporting References}

[S1]
\textrm{Bernevig, B. A.}, \textrm{Hughes, T. L.} \& \textrm{Zhang, S. C.}, \textrm{Quantum spin Hall effect and topological
phase transition in HgTe quantum wells}. {\it Science\/} {\bf 314},
 1757-1761 (2006).

\noindent
[S2]
 \textrm{Juri\v{c}i\'{c}, V.},  \textrm{Mesaros, A.}, \textrm{Slager, R.-J.} \& \textrm{Zaanen, J.}, \textrm{Universal Probes of Two-Dimensional Topological Insulators:
 Dislocation and $\pi$ Flux}.
 {\it Phys. Rev. Lett.\/} {\bf 108}, 106403 (2012).

\noindent
 [S3]
 \textrm{Kane, C. L.} \& \textrm{Mele, E. J.}, \textrm{$Z_{2}$ Topological Order and the Quantum Spin Hall Effect}. {\it Phys. Rev. Lett.\/} {\bf 95},
  146802 (2005).

\noindent
 [S4]
  \textrm{Fu, L.} \& \textrm{C. L. Kane}, \textrm{Topological insulators with inversion symmetry}, {\it Phys. Rev. B\/} {\bf 76},
  045302 (2007).

\noindent
[S5]
\textrm{Murakami, S.}, \textrm{Quantum Spin Hall Phases}. {\it Prog. Theor. Phys. Suppl. \/} {\bf 176},
 279-301 (2008).

\noindent
[S6]
\textrm{Murakami, S.} \& \textrm{Kuga, S.},  \textrm{Universal phase diagrams for the quantum spin Hall systems}. {\it Phys. Rev. B\/} {\bf 78},
165313 (2008).

\noindent
[S7]
\textrm{Lu, C.-K. \& Herbut, I. F.}, \textrm{Zero modes and charged Skyrmions in graphene bilayer}. {\it Phys. Rev. Lett.\/} {\bf 108}, 266402 (2012) .


\newpage

\begin{thebibliography}{10}

\bibitem{hasan-kane-review}
\textrm{M.\ Z.\ Hasan}, \textrm{C.\ L.\ Kane}, \textrm{Colloquium: Topological insulators}. {\it Rev. Mod. Phys.} {\bf 82}, 3045 (2010).

\bibitem{qi-zhang-review}
\textrm{X.\ L.\ Qi}, \textrm{S.\ C.\ Zhang}, \textrm{Topological insulators and superconductors}. {\it Rev. Mod. Phys.} {\bf 83}, 1057 (2011).

\bibitem{tknn1982}
\textrm{D. J. Thouless}, \textrm{M. Kohmoto},\textrm{ M. P. Nightingale}, \textrm{M. den Nijs}, \textrm{ Quantized Hall Conductance in a Two-Dimensional Periodic
Potential}.
 {\it Phys. Rev. Lett.\/} {\bf 49}, 405 (1982).

 \bibitem{kane2005}
\textrm{C. L. Kane}, \textrm{E. J. Mele}, \textrm{$Z_{2}$ Topological Order and the Quantum Spin Hall Effect}. {\it Phys. Rev. Lett.\/} {\bf 95},
  146802 (2005).

  \bibitem{moore2007}
\textrm{J.E. Moore}, \textrm{L. Belents}, \textrm{Topological invariants of time-reversal-invariant band structures}. {\it Phys. Rev. B\/} {\bf 75},
121306 (2007).

\bibitem{fukane2006}
\textrm{L. Fu}, \textrm{C. L. Kane}, \textrm{Time reversal polarization and a $Z_{2}$ adiabatic spin pump}. {\it Phys. Rev. B\/} {\bf74},
 195312 (2006).

 \bibitem{fukane2007a}
\textrm{L. Fu}, \textrm{C. L. Kane}, \textrm{Topological Insulators in Three Dimensions}. {\it Phys. Rev. Lett.\/} {\bf 98},
  106803 (2007).

\bibitem{fukaneMajorana2008}
  \textrm{L. Fu}, \textrm{C. L. Kane}, \textrm{Superconducting Proximity Effect and Majorana Fermions at the Surface of a Topological Insulator}.
  {\it Phys. Rev. Lett.\/} {\bf 100}, 096407 (2008).

  \bibitem{Zhang-Theta2010}
  \textrm{R. D. Li, J. Wang, X. L. Qi, S. C. Zhang}, \textrm{Dynamical axion field in topological magnetic insulators}, Nat. Phys. {\bf 6}, 284 (2010).



 \bibitem{bernevigzhang2006}
\textrm{B. A. Bernevig}, \textrm{T.L. Hughes}, \textrm{S.C. Zhang}, \textrm{Quantum spin Hall effect and topological
phase transition in HgTe quantum wells}. {\it Science\/} {\bf 314},
 1757 (2006).

\bibitem{zhang2009}
H.~ Zhang, {\it et~al.\/}, \textrm{Topological insulators in $\text{Bi}_{2}\text{Se}_{3}$, $\text{Bi}_{2}\text{Te}_{3}$ and $\text{Sb}_{2}\text{Te}_{3}$ with a
single Dirac cone
on the surface}. {\it Nature Phys.\/} {\bf 5}, 438 (2009).

 \bibitem{koning2006}
 M.~Koning, {\it et~al.\/}, \textrm{Quantum spin Hall insulator state in HgTe quantum wells}. {\it Science\/} {\bf 318}, 766 (2007).

\bibitem{hsieh2008}
 D.~Hsieh, {\it et~al.\/}, \textrm{A topological Dirac insulator in a quantum spin Hall phase}. {\it Nature\/} {\bf 452}, 970 (2008).

\bibitem{hsieh2009}
 D.~Hsieh, {\it et~al.\/}, \textrm{Observation of unconventional quantum spin textures in topological
insulators}. {\it Science\/} {\bf 323}, 919 (2009).

 \bibitem{xia2009}
Y.~ Xia, {\it et~al.\/}, \textrm{Observation of a large-gap topological-insulator class with a single Dirac cone
on the surface}. {\it Nature Phys.\/} {\bf 5}, 398 (2009).

\bibitem{chen2009}
Y. L. ~Chen, {\it et~al.\/}, \textrm{Experimental realization of a three-dimensional topological insulator,
$\text{Bi}_{2}\text{Te}_{3}$}. {\it Science\/} {\bf 325}, 178 (2009).

  \bibitem{schnyder2008}
\textrm{A. P. Schnyder}, \textrm{ S. Ryu},\textrm{  A. Furusaki}, \textrm{A.W.W. Ludwig}, \textrm{ Classification of topological insulators and superconductors in
three spatial dimensions}.
{\it Phys. Rev. B\/} {\bf 78}, 195125 (2008).

\bibitem{schnyderNJP2010}
\textrm{S. Ryu, A. P. Schnyder, A. Furusaki, A. Ludwig}, \textrm{Topological insulators and superconductors: ten-fold way and dimensional hierarchy}.
{\it New J. Phys.}, {\bf 12}, 065010 (2010).

 \bibitem{kitaev2009}
 \textrm{A. Kitaev}, \textrm{Periodic table for topological insulators and superconductors}.  {\it AIP Conf. Proc.\/} {\bf 22},
  1132 (2009).

\bibitem{ranNatPhys2009}
 \textrm{Y. Ran}, \textrm{Y. Zhang}, \textrm{A. Vishwanath}, \textrm{One-dimensional topologically protected modes in
topological insulators with lattice dislocations}. {\it Nat. Phys.\/} {\bf 5},  298 (2009)

\bibitem{fukane2007b}
\textrm{L. Fu}, \textrm{C. L. Kane}, \textrm{Topological insulators with inversion symmetry}, {\it Phys. Rev. B\/} {\bf 76},
  045302 (2007).

  \bibitem{teofukane2008}
  \textrm{J. C. Y. Teo, L. Fu, C. L. Kane}, \textrm{Surface states and topological invariants in three-dimensional topological insulators:
Application to $\text{Bi}_{1-x}\text{Sb}_x$}, {\it Phys. Rev. B}, {\bf 78}, 045426 (2008).

  \bibitem{fu2011}
\textrm{L. Fu}, \textrm{Topological Crystalline Insulators}. {\it Phys. Rev. Lett.\/} {\bf 106},
106802 (2011).

\bibitem{fuNatComm2012}
\textrm{T.\ H.\ Hsieh, H. Lin, J. Liu, W. Duan, A. Bansil, L. Fu}, \textrm{ Topological Crystalline Insulators in the SnTe Material Class}.
{\it arXiv:1202.1033v2} (to appear in Nat. Comm.) (2012).

\bibitem{hughesprodanbernevigPRB2011}
\textrm{T. L. Hughes, E. Prodan, B. A. Bernevig}, \textrm{Inversion-symmetric topological insulators}. {\it Phys.\ Rev.\ B} {\bf 83}, 245132 (2011).

\bibitem{turnerPRB2012}
\textrm{A. M. Turner, Y. Zhang, R. S. K. Mong, A. Vishwanath}, \textrm{Quantized Response and Topology of magnetic insulators with inversion symmetry}.
{\it Phys.\ Rev.\ B } {\bf 85}, 165120 (2012).

 \bibitem{juricic2012}
 \textrm{V. Juri\v{c}i\'{c}},  \textrm{A. Mesaros}, \textrm{R.J. Slager}, \textrm{J. Zaanen}, \textrm{Universal Probes of Two-Dimensional Topological Insulators:
 Dislocation and $\pi$ Flux}.
 {\it Phys. Rev. Lett.\/} {\bf 108}, 106403 (2012).

\bibitem{kane2005a}
\textrm{C.\ L.\ Kane, E. J. Mele}, \textrm{Quantum Spin Hall Effect in Graphene}. {\it Phys. Rev. Lett.} {\bf 95}, 226801 (2005).

\bibitem{zhangfelser2012}
\textrm{X. Zhang, H. Zhang, J. Wang, C. Felser, S. C. Zhang}, \textrm{Actinide Topological Insulator Materials with Strong Interaction }.
{\it Science}, {\bf 335}, 1464 (2012).

\bibitem{us-inpreparation}
R.-J. Slager, A. Mesaros, V. Juri\v ci\' c, J. Zaanen, {\it in preparation}.

\bibitem{ando2012}
Y.~ Tanaka {\it et al.}, \textrm{Experimental realization of a topological crystalline insulator in $\text{SnTe}$}. {\it arXiv:1206.5399v1} (2012).

\bibitem{xu2012}
S.Y.~ Xu {\it et~al.\/}, \textrm{Observation of Topological Crystalline Insulator phase in the lead tin chalcogenide $\text{Pb}_{1-x}\text{Sn}_{x}\text{Te}$
material class}. {\it arXiv:1206.2088v1} (2012).

\bibitem{story2012}
P.~ Dziawa {\it et al.}, \textrm{Topological crystalline insulator states in $\text{Pb}_{1-x}\text{Sn}_x\text{Se}$}. {\it arXiv:1206.1705v1}
(advanced online publication in Nature Materials) (2012).

 \bibitem{dresselhausbook}
 \textrm{M. S. Dresselhaus, G. Dresselhaus, A. Jorio}, {\it Group Theory Application to the Physics of Condensed Matter} (Springer, Berlin, 2008).



\end{thebibliography}
\end{document}